\documentstyle[aps,eqsecnum,preprint,psfig]{revtex}
\begin{document}
\draft
\title{Collapsed and adsorbed states of a directed polymer chain in 
two-dimensions.}
\author{Pramod K Mishra and Yashwant Singh }
\address{ Department of Physics, Banaras Hindu University, \\
     Varanasi 221 005, India }
\date{\today}
\maketitle
\begin{abstract}
A phase diagram for a surface interacting long flexible partially
directed polymer chain in a two-dimensional poor solvent where the 
possibility of collapse in the bulk exists is determined using exact 
enumeration method. We used a model of self attracting self-avoiding 
walk and evaluated $30$ steps in series. An intermediate phase in 
between the desorbed collapsed and adsorbed expanded phases having 
the conformation of a surface attached globule is found. The four 
phases, viz. $ (i)$ desorbed expanded, $(ii)$ desorbed collapsed, 
$(iii)$ adsorbed expanded, $(iv)$ surface attached globule are found 
to meet at a multicritical point. These features are in agreement 
with those of an isotropic (or non directed) polymer chain.
\end{abstract}                                                               

\narrowtext 

\section{Introduction}
The subject of adsorption of a long flexible polymer chain immersed 
in a poor solvent on an impenetrable surface with an attractive 
short-range attraction has received considerable attention in recent 
years \cite{1,2,3,4}. This is because an attractive surface may lead 
to adsorption-desorption transition from the state when the chain is 
mostly attached to the surface to the state of detachment when the
temperature is increased. This behaviour finds applications in 
lubrication, adhesion, surface protection, etc. 

It is known that when a chain interacts with a surface its 
conformational properties get modified in comparison with its bulk 
properties \cite{1,2}. This is because of a subtle competition 
between the gain of internal energy and the corresponding loss of 
entropy due to the constraint imposed on the chain by the impenetrable 
surface. This competition leads to the possibility of co-existence 
of different regimes and multicritical behaviour.

The essential physics associated with the behaviour of a surface
interacting polymer chain in a good solvent  is derived
from a model of self-avoiding walk (SAW) on a semi-infinite lattice.
If the surface is attractive, it contributes an energy $\epsilon_s$
($< 0$) for each step of the walk on the surface. This leads
to an increased probability characterized by the Boltzmann factor
$\omega = \exp(-\epsilon_s/k_{\beta}T)$ of making a step along the surface,
since for $\epsilon_s < 0$, $\omega > 1$ for any finite temperature $T$
($k_{\beta}$ is the Boltzmann constant). Because of this the polymer
chain get adsorbed at low temperatures on the surface while at
high temperatures all polymer conformations have almost same weight
and non adsorbed (or desorbed) behaviour prevails. The transition
between these two regimes is marked by a critical adsorption
temperature $T_a$, with a desorbed phase for $T > T_a$ and adsorbed
phase for $T < T_a$ as shown in Fig. $1$. At $T < T_a$ the chain acquires
a conformation such that a fraction of monomers get attached to the 
surface and others form a layer of finite thickness parallel to the 
surface.  The thickness of the layer diverges at $T=T_a$. One may 
define the crossover exponent $\phi$, at $T=T_a$ as
$N_s \sim N^{\phi}$ where $N$ is the total number of monomers
and $N_s$ the number of monomers on the surface. The transition point
$T_a$ is a tricritical point \cite{2}. Both the surface and the bulk
critical exponents have been calculated using renormalization group
methods \cite {5}, transfer matrix methods \cite{6,7},
exact enumeration method \cite {8} and Monte Carlo methods \cite {9}. 
For a two dimensional system exact values of the exponents
have been found by using conformal invariance \cite {10}.

\vspace {.4in}

\psfig{figure=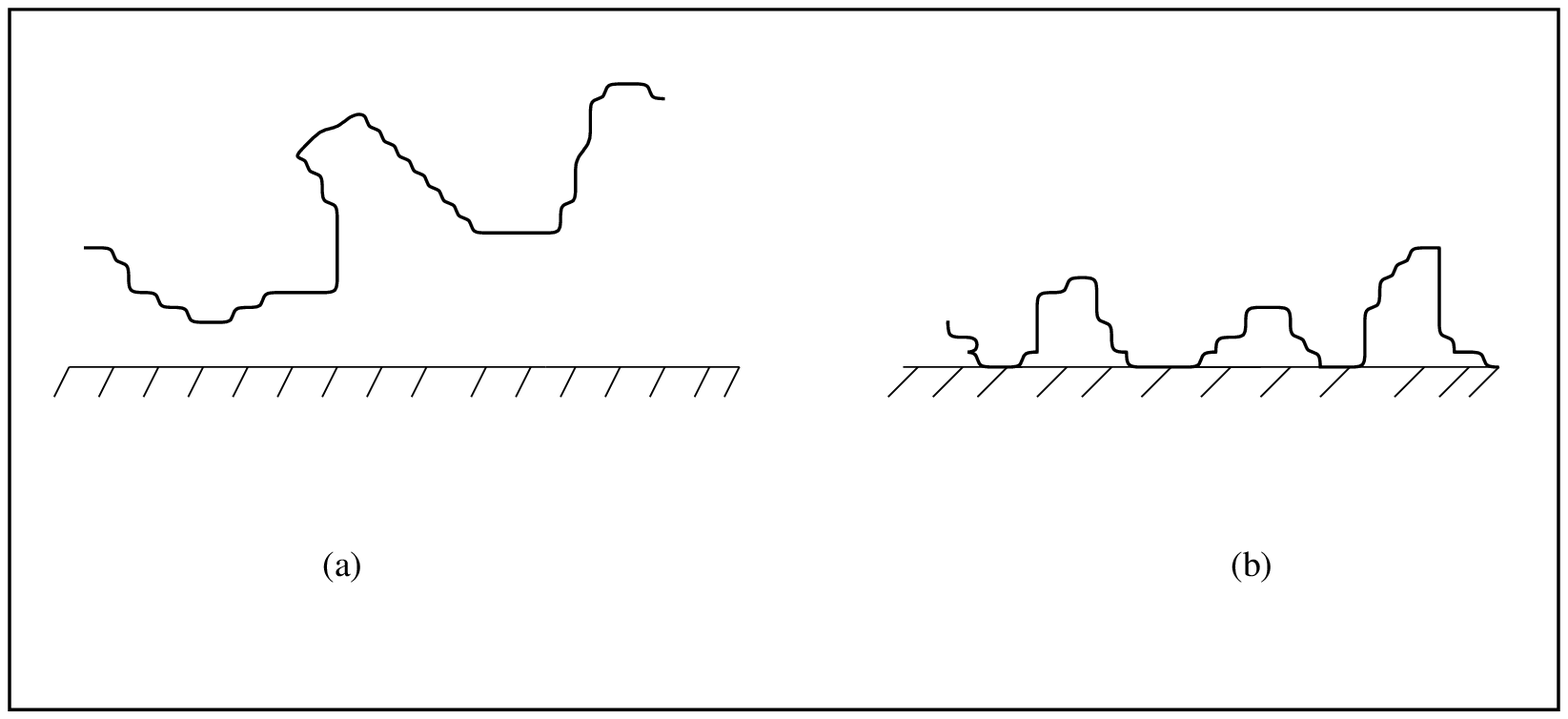,height=3in,width=6.6in}
\vspace {.1in}
{\bf Fig. 1; \small Schematic representation of polymer conformations in 
(a) desorbed and (b) adsorbed states.}
\vspace {.2in}

The situation is, however, different when the surface interacting 
polymer is a in compact globule state. In the vicinity of an 
impenetrable attracting surface the monomer-monomer attraction which 
is responsible for the collapsed state of the chain and the 
surface-monomer attraction responsible for the adsorption
will compete. This competition may give rise new features in the 
conformational behaviour of the chain. For instance, a globule can 
get attached to the surface at some value of monomer-surface 
attraction without any significant modification in its structure as 
shown in Fig. $2$. Such a phase has been shown to exist in between
the bulk collapsed and adsorbed phases in both two and three dimensions 
for isotropic ${i. e.}$ non-directed) polymer chain and
is called surface attached globule (SAG) phase \cite{11,12}. In an 
adsorbed phase the number of monomers on the surface are proportional 
to $N$ and the thickness of the layer formed by the chain
parallel to the surface proportional to $N^0$. On the other hand the 
bulk globule has one correlation length (the radius of gyration) which 
scales with $N$ as  $R \propto N^{1/d}$ where d is the dimensionality
of the space. In the SAG phase we have two correlation lengths, one 
parallel and another perpendicular to the surface. They may scale with 
$N$ as $R_{\parallel} \propto N^{{x}_{\parallel}}$ and $R_{\perp} 
\propto N^{{x}_{\perp}}$ where $\frac {1}{d} \le {x}_{\parallel} \le 
\frac {1}{d-1}$ and $0 < {x}_{\perp} \le \frac {1} {d}$.

\vspace {.4in}

\psfig{figure=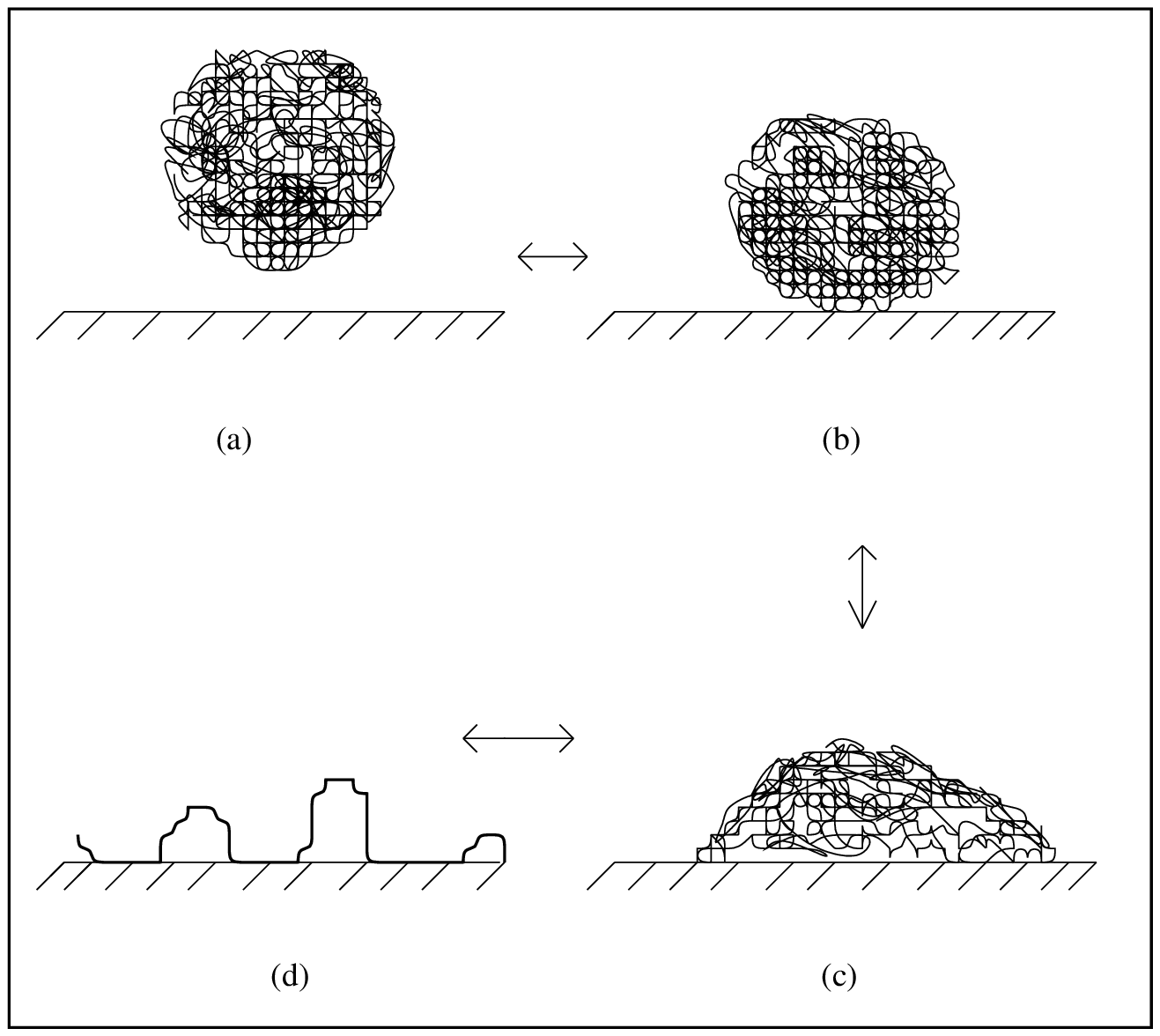,height=4in,width=5.6in}
\vspace {.2in}
{\bf Fig. 2; \small Schematic representation of adsorption of a compact 
globule; (a) shows a desorbed globule, (b) and (c) surface attached 
globule and (d) the adsorbed state.}
\vspace {.4in}

In this article we investigate the surface adsorption and collapsed 
of a directed polymer chain on a square lattice. Since for two 
dimensional system surface is a line, it can have only one adsorbed 
phase which we refer to as adsorbed expanded (AE) phase. The problem 
of adsorption and collapse of directed polymer has been studied 
using transfer matrix technique \cite {13}. The phase boundary of 
the collapsed phase separating it from the adsorbed expanded and 
the bulk expanded has been obtained exactly \cite {14} as well as 
numerically \cite {15}. These calculations show that there are three 
phases, viz. desorbed expanded (DE), desorbed collapsed (DC) and 
adsorbed expanded (AE). These three phases meet at a multicritical 
point. In these calculations a condition $xu = 1$ (where $x$ is the 
step fugacity and $u$ the Boltzmann factor for the monomer-monomer 
attraction) is used. This condition is valid only at the collapsed 
phase boundary. Therefore only the phase boundaries separating the 
DC phase from the AE and DE phases were obtained by these calculations. 
The phase boundary separating the DE phase from the AE phase has been 
obtained numerically by Veal ${\it et\; al}$ \cite {15}.  Since 
the transition from the DC phase to the SAG phase takes place in the 
region where $xu > 1$ the calculations based on the transfer matrix
could not show this transition.  

\section { Calculational details }
We consider partially directed self-attracting self-avoiding 
walks (PDSASAWs) on a square 
lattice restricted to quarter space. Walk starts from the corner 
of the surface. Let $C_{N,N_s,N_p}$ be the number of PDSAWs with
$N$ steps, having $N_s$ $(\le N)$ steps on the surface and $N_p$
nearest neighbours. We have obtained $C_{N,N_s,N_p}$ for $N \le 30$ 
for square lattice by exact enumeration method. We prefer this 
technique because in this case the scaling corrections are 
correctly taken into account by a suitable extrapolation scheme \cite{9}. 
 
We associate energy $\epsilon_s$ with each monomer on the surface 
and $\epsilon_p$ for the monomer-monomer interaction. Partition 
function of the attached chain is 
\begin{equation}
Z_N (\omega, u) = \sum_{N_s,N_p} C_{N,N_s,N_p} \omega^{N_s}
u^{N_p} 
\end{equation}
where $\omega = e^{-\epsilon_s/kT}$ and $u = e^{- \epsilon_p/kT}$.
$\omega > 1$ and $u >1$ for attractive force. Reduced free
energy for the chain can be written as 
\begin{equation}
G (\omega, u) = \lim_{N\rightarrow \infty} \frac{1}{N} \log
Z_N(\omega,u) 
\end{equation}
In general it is appropriate to assume that as $N\rightarrow \infty$ 
\begin{equation}
Z_{N}(\omega,u) \sim N^{\gamma - 1} \mu(\omega,u)^{N} 
\end{equation}
where $\mu(\omega,u)$ is the effective coordination number and
$\gamma$ is the universal configurational exponents for walks
with one end attached to the surface. The value of
$\mu(\omega,u)$ can be estimated using ratio method \cite {16}
with associated Neville table or any other method such as Pade
analysis \cite {17} or differential approximants \cite{18,19}. 

From equations (2.2) and (2.3) we can write 
\begin{equation}
\log \mu(\omega, u) = \lim_{N\rightarrow \infty} \frac{1}{N} \log
Z_N(\omega,u) = G (\omega, u)
\end{equation}

$Z_N(\omega,u)$ is calculated from the data of $C_{N,N_s,N_p}$
using equation (2.1) for a given $\omega$ and $u$. From this we
construct linear and quadratic extrapolants of the ratio of
$Z_N(\omega,u)$ for the adjacent values of $N$ as well as the
alternate one. Results for alternate $N$ give better
convergence. When $u = 1$ and $\omega = 1$ the value of $\mu$ is
found to be 2.39.

The surface critical exponent $\gamma_1$ can be calculated using 
the relation \cite {8}
\begin{equation}
\gamma^0 - \gamma_1 = \frac{\log(Z_N^0 Z_{N-2} / Z_{N-2}^0
Z_N)}{\log(N/N-2)}
\end{equation}
and the end-to-end distance exponent $\nu$ using the relation 
\begin{equation}
\nu = \frac{1}{2} \frac{\log(R_N^2 / R_{N-2}^2)}{\log(N/N-2)}
\end{equation}
where the superscript ``0" indicates that the  quantities 
correspond to the bulk without the surface and $R_N$ is 
the end-to-end distance of a chain of $N$ monomers.

The value of $\omega_c (u)$ at which polymer gets adsorbed/attached 
to the surface for a given value of $u$ is found from the $(i)$ plot 
of $G_s(\omega,u)=G(\omega,u)-G(u,\omega=1)$ 
which remains equal to zero until $\omega = \omega_c$ and 
increases consistently as a function of $\omega$ for 
$\omega \ge \omega_c, $ $(ii)$ from the plot of 
$\partial^2 G_s(\omega,u) / \partial {\epsilon_s}^2$ at constant
$u$  and 
$(iii)$ from the plot of $\gamma^0 - \gamma_1$ as a function of 
$\omega$ for different $N$. The value of $\omega_c$ found from the 
plot of $G_s(\omega,u)$ is slightly lower than the peak value of 
$\partial^2 G_s/ \partial {\epsilon_s}^2$. It is, however, observed
that as $N$ is increased from $22$ to $30$  the peak value shifts 
to smaller $\omega$ and appears to converge on the value of 
$\omega$ found from $G_s(\omega,u)$ plot.
We therefore choose the value of $\omega_c$ found from the plot 
of $G_s(\omega,u) $ and determine lines $\omega_c(u)$ and 
$\omega_{c1} (u)$ (see Fig. 3) by this method.
For $u = 1$, the value of $\omega_c$ is = $1.70$ which is in very
good agreement with the value $1.707$ reported in ref.
\cite {6}. In another approach equation (2.5) is used to calculate 
$\gamma^0 - \gamma_1$ for different $N$ and the values are plotted 
as a function of $\omega$. The value of $\omega_c$ is found from
the intersection of successive approximation to $\gamma^0 - \gamma_1$
in the limit $N \rightarrow \infty$. This method, however, fails 
for $u$ values above the $\theta$-point, {\it i. e.} where the 
desorbed phase is in a collapsed globule state. Below the $\theta$-point 
both the methods yield same values of $\omega_c$ which are in very good 
agreement.

The phase boundary separating the expanded and  collapsed phases is 
calculated from the plot of $G_b(\omega,u)=G(\omega,u)-G(\omega,u=1)$ 
as a function of $u$ for a given $\omega$. Here $G_b$ measures the 
energy arising due to monomer-monomer attractions. This is however, 
not zero in the expanded phase and, therefore, cannot be used as an 
order parameter for the expanded and collapsed transition. We, however, 
find that the value of $G_b(\omega,u)$ as a function of $u$ shows 
sudden rise at certain value of $u$. We locate this point from the peak 
of $\partial^2 G_b/\partial \epsilon_p^2$ which corresponds to a specific 
heat peak and shows divergence as $N \rightarrow \infty$. 
For $\omega = 1$, the value of $u_c$ is 3.30 which is in good agreement
with the value 3.38 found by exact calculation. The method is found 
to work for all values of $\omega$ {\it i.e} in both the bulk and the 
adsorbed regimes. However, as $\omega$ is increased the values of 
$G_b (\omega,u)$ do not remain as smooth as at lower values of $\omega$, 
therefore introducing some inaccuracy in the values of $u_c$.
The values of $(\omega_c,u_c)$ at the multicritical point found by us 
are $(2.16,3.30)$ which are in good agreement with exactly known values 
$(2.19,3.38)$. As $\omega$ and $u$ becomes larger, higher order terms 
in equation (2.1) become significant and therefore to have proper 
convergence more terms must be considered. This explains the difference 
in $\omega_{c2}$ line found by exact calculation \cite {14} and the 
present calculation (see fig. 3). 

\section{Results}
The phase diagram shown in Fig. 3 has four phase boundaries instead
of three as reported in earlier work \cite{14,15}. The line $u_c$ 
 separates the expanded and collapsed phases. 

\vspace {.3in}

\psfig{figure=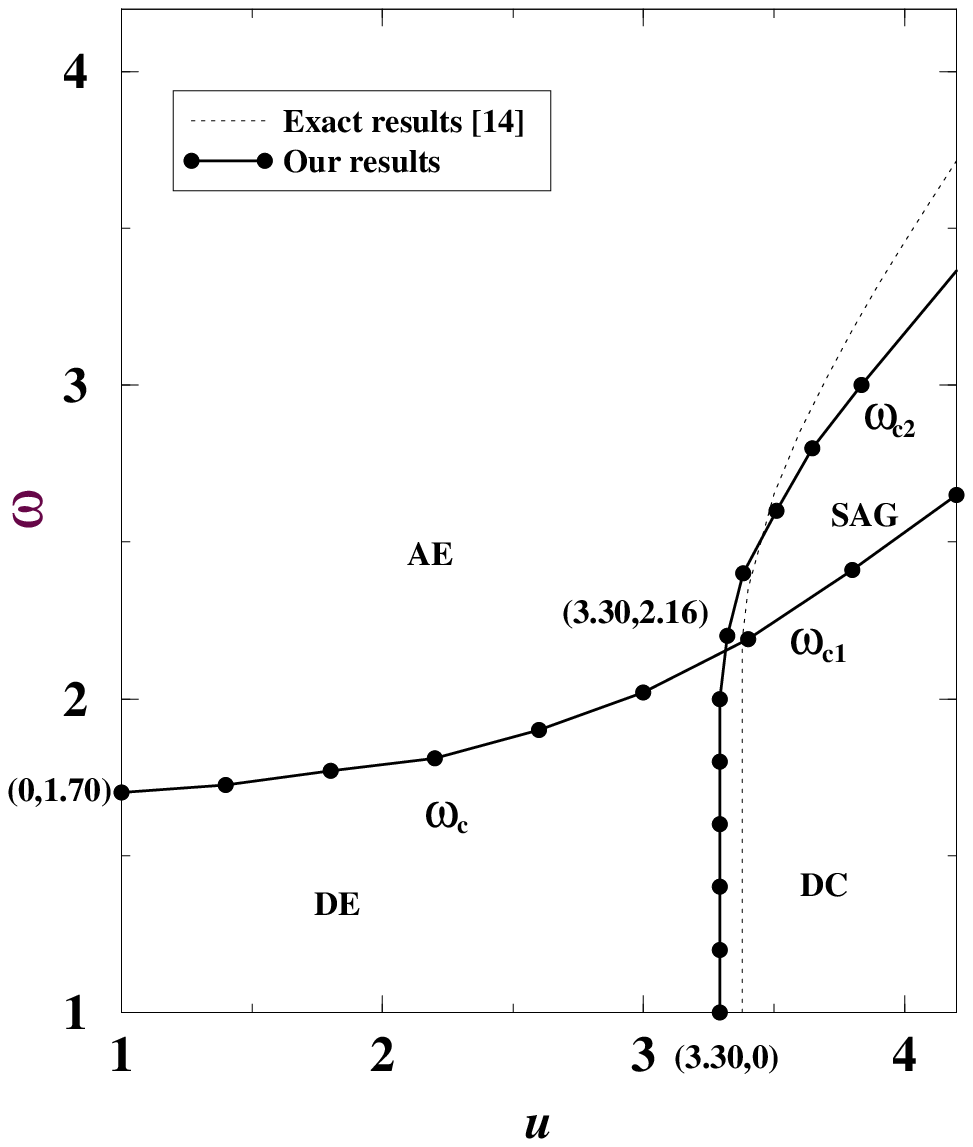,height=4.0in,width=4.0in}

\vspace {.1in}

{\bf Fig. 3; \small Phase digram of a directed linear 
polymer chain in 2-D space. The $\omega$ and $u$ axes represents, the 
Boltzmann factor of surface attraction and monomer-monomer attraction 
respectively. Solid line curve corresponds to present calculation and
the dotted line to ref. \cite{14}.}

\vspace {.3in}

The special adsorption line $\omega_c$ separates adsorbed expanded (AE) 
phase from that of desorbed expanded (DE). Beyond $\theta$-point we have 
two boundaries $\omega_{c1}$ and $\omega_{c2}$. The line $\omega_{c1}$ 
separates the desorbed collapsed (DC) bulk phase from that of a surface 
attached globule (SAG) state, whereas the boundary $\omega_{c2}$ separates 
the SAG phase from the AE phase. The point where $u_c$ line meets the 
special adsorption line $\omega_c$, all four phases AE, DE, DC and SAG 
coexist. The SAG phase which exists between the boundaries $\omega_{c1}$ 
and $\omega_{c2}$ for $u > u_c$ is essentially a two-dimensional globule 
sticking to the surface in the same way as a liquid droplet may lie on 
a partially wet surface. These features are in agreement with those of 
isotropic (non directed) polymer chain. The average fraction of monomer 
on the surface and the average number of pairs are calculated using the 
relations.
\begin{displaymath}
<n_s> = \lim_{N\rightarrow\infty} w\frac{\partial G}{\partial\omega}
\mid_{u}, \;\; \; 
and
<n_p> = \lim_{N\rightarrow\infty} u\frac{\partial G}{\partial u}
\mid_{\omega} 
\end{displaymath}

The results are shown in Figs. $4$ and $5$. The transition points on 
each curve is marked by dots. The dot on a curve of $<n_s>$ (in Fig. 4) 
indicates adsorption or polymer chain getting attached to the surface 
({\it i.e} $\omega_c$ and $\omega_{c1}$ lines depending on the values 
of $u$), the dot on a curve $<n_p>$ (in Fig. 5) indicates the 
collapsed-expanded states transition ({\it i. e.} point on line $u_c$ 
and $\omega_{c2}$ depending on the values of $\omega$). We may note 
that the value of $<n_p>$ for SAG phase is comparable to that in the 
DC phase. 

\vspace {.3in}

\psfig{figure=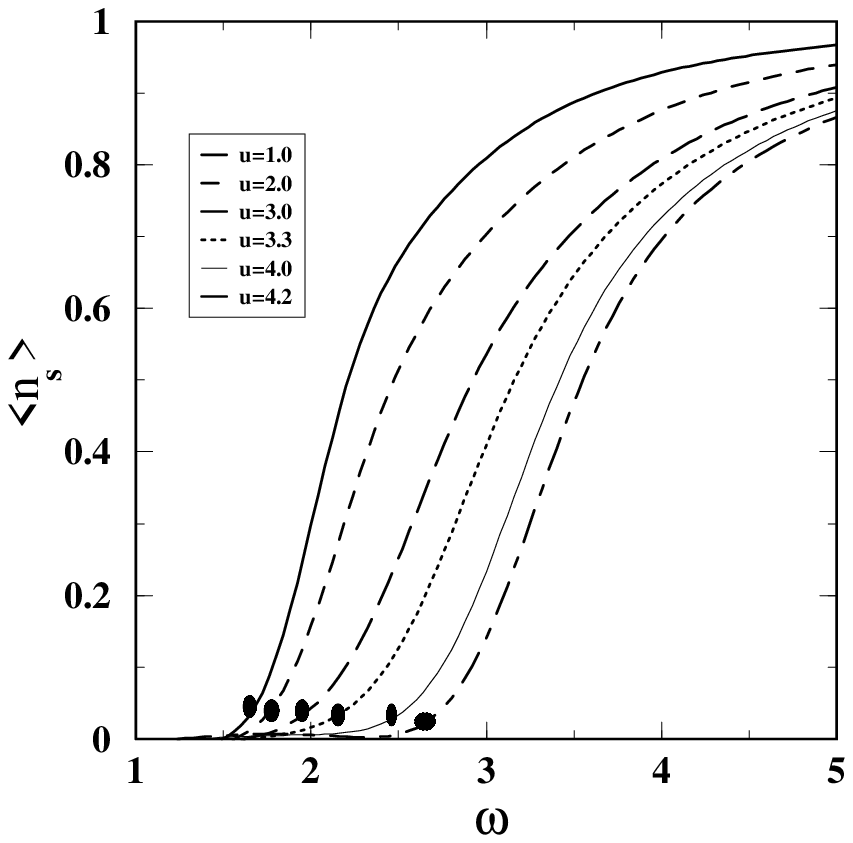,height=4.0in,width=4.0in}

\vspace {.1in}

{\bf Fig.4; \small Average fraction of monomers on the surface $<n_s>$
as a function of $\omega$. Dot on the curve represents transition 
point.}

\vspace {.2in}

We have found that along the line $\omega_c$ and $\omega_{c1}$, 
$<n_s> = 0.04 \pm 0.01$ and along
the line $u_c$ and $\omega_{c2}$, $<n_p> = 0.54 \pm 0.06$.

\vspace {.3in}

\psfig{figure=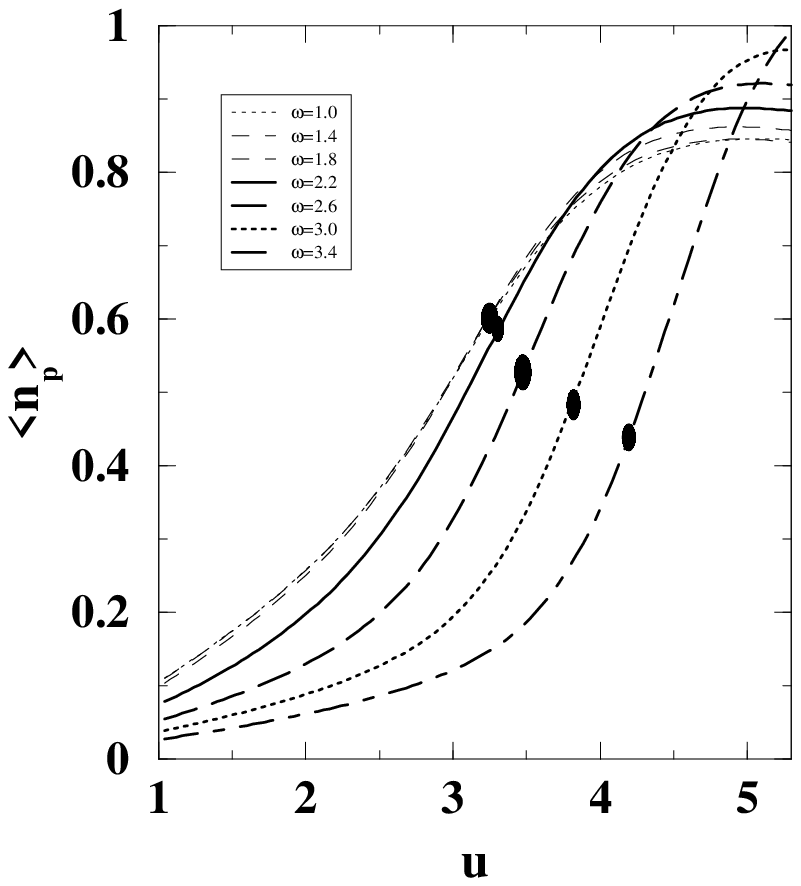,height=4.0in,width=4.0in}

\vspace {.1in}

{\bf Fig.5; \small Average number of pairs $<n_p>$ as a function of $u$.
Dot on the curve represents the transition point.}

\vspace {.2in}

\section{Conclusions}
It is obvious from these results that when the chain gets adsorbed 
from the expanded bulk state ({\it i.e} for $u < u_c$), it acquires 
a conformation at $\omega \simeq \omega_c (u)$ such that a small 
fraction of monomers get attached to the surface  and others forms 
a layer parallel to the surface. This affects the free energy of 
the system. At the transition point the polymer chain fluctuates 
between adsorbed and desorbed conformations. Therefore one may get 
peak in $<N_s^2>-<N_s>^2$ at $\omega=\omega_c$ which shows divergence 
as $N \rightarrow \infty$. As $\omega$ is increased for the same 
value of $u$, the fluctuations along the normal to the surface get 
suppressed and at large $\omega (>> \omega_c)$ the chain lies on 
the surface with very little fluctuations. On the other hand, 
when the adsorbing chain was in collapsed bulk state, then at 
$\omega = \omega_{c1}$ the collapsed chain gets attached to the 
surface. Here again the number of monomers getting attached to the 
surface are small. Since for $\omega \leq \omega_{c1}$ the globule 
may be any where in the available space and get attached to the surface 
at $\omega=\omega_{c1}$, there is a breaking of translational invariance 
at $\omega=\omega_{c1}$ For $\omega_{c1} \leq \omega \leq \omega_{c2} (u)$
the chain remains in the form of a globule attached to the surface. 
In this range the monomer-monomer attraction remains effective in holding 
the monomers in the neighbourhood of each other than the surface-monomer 
attraction whose tendency is to spread the chain on the surface. Since 
at $\omega=\omega_{c1}$ the globule retains its shape there is no change
in the bulk free energy. Only the surface contribution to the free energy
which was zero for $\omega\leq\omega_{c1}$ starts contributing. 
Therefore, in this case we may therefore not see divergence in the bulk 
specific heat peak in the $N \rightarrow \infty$ limit. For $\omega > 
\omega_{c2} (u)$ the globule conformation becomes unstable as 
surface-monomer attraction becomes more effective than the 
monomer-monomer attraction and therefore the chain spreads over the 
surface (just like a liquid droplet spreads over a wetting surface).

Summarizing we studied a polymer chain immersed in a poor solvent 
in the presence of an attracting impenetrable wall and obtained the 
phase boundaries separating different phases of the polymer chain 
from data obtained by exact enumerations. We report a state in 
between the desorbed collapsed and adsorbed phases which has 
conformation of a compact globule sticking to a surface in same 
way as a droplet may lie on a partially wet surface. A similar 
phase also been found in case of isotropic (non directed) polymer 
chain in both two and three dimensions \cite {11,12}. The  
number of monomers on the surface and the number of nearest 
neighbors in different regimes of the phase diagram are obtained.  

\section { Acknowledgements }

We are thankful to Dr. S. Kumar and Dr. D. Giri for many 
fruitful discussions. The work was supported by the Department of
Science and Technology (India) through project grant.

\end{document}